\documentclass[12pt]{book}
\usepackage{epsfig}
\usepackage{graphicx}
\usepackage{pazhe}
\tightenlines
\def\deg{^\circ}

\begin{document}

ISSN 1063-7737, Astronomy Letters, 2007, Vol.33, No. 12, pp. 807-813. \copyright Pleiades Publishing, Inc., 2007.

\title
{\bf REGISTRATION OF THE FIRST THERMONUCLEAR X-RAY BURST FROM AX\,J1754.2-2754}

\author
{\bf \hspace{-1.3cm}\copyright\, 2007 Ç. \ \ 
I.V. Chelovekov\affilmark{1*}, S.A. Grebenev\affilmark{1}}

\affil{{\it Space Research Institute RAN, Moscow}$^1$\\}
\vspace{2mm}
\received{June 14, 2007}
\vspace{2mm}
\baselineskip 15pt
\noindent
During the analysis of the INTEGRAL observatory archival data
we found a powerful X-ray burst, registered by JEM-X and IBIS/ISGRI
telescopes on April 16, 2005 from a weak and poorly known source
AX\,J1754.2-2754. Analysis of the burst profiles and spectrum shows, 
that it was a type I burst, which result from thermonuclear explosion
on the surface of nutron star. It means that we can consider 
AX\,J1754.2-2754 as an X-ray burster. Certain features of burst profile
at its initial stage witness of a radiation presure driven strong 
expansion and a corresponding cooling of the nutron star photosphere.
Assuming, that the luminosity of the source at this phase was close 
to the Eddington limit, we estimated the distance to the burst 
source $d=6.6\pm0.3$ kpc (for hidrogen atmosphere of the neutron star)
and $d=9.2\pm0.4$ kpc (for helium atmosphere). 


\vfill

\noindent\rule{8cm}{1pt}\\
{$^*$ e-mail: $<$chelovekov@hea.iki.rssi.ru
\vspace{3cm}
$>$}

\clearpage

\section*{INTRODUCTION}

X-ray burster AX J1754.2-2754 was first observed on October 2--3,
1999 by ASCA (Sakano et al., 2002) orbital observatory. Its 0.7--10 
keV flux reached $\sim10^{-11}$ erg cm$^{-2}$ s$^{-1}$ ($\sim0.3$ mCrab). 
The source had a 
power law spectrum with an index $\alpha\sim2.5$, modified
at low energies by interstellar absorption corresponding to $N_{\rm
H}\sim2\times10^{22}$ cm$^{-2}$. Types of the source and its 
compact object were not determened at that time. AX J1754.2-2754
was registered (at the IBIS/ISGRI sensitivity limit --- with $2.05\pm0.53$ 
mCrab in 17-60 keV energy band) later during a hard X-ray 
INTEGRAL observatory 2003-2006 all-sky servey (Krivonos et al., 2007). 
These observations as well failed to discover its true nature.

In this paper we report the discovery of type I X-ray burst from 
this source with INTEGRAL observatory telescopes (see also
Chelovekov, Grebenev, 2007) and present the results of its detailed
analysis. Registration of this burst puts AX J1754.2-2754 in a group
of X-ray bursters, so the compact object in this source is a neutron 
star.

\section*{INSTRUMENTS AND OBSERVATIONS}

INTEGRAL (Winkler et al., 2003) is an international orbital 
$\gamma$-observatory, that was put to its high apogee orbit 
by the PROTON launcher on October 17, 2002 (Eismont et al, 2003). 
The observatory consists of four telescopes, meant to perform 
simultaneous observations in $\gamma$-ray, X-ray and optical 
energy bands. In this paper we use data from JEM-X X-ray monitor
(Lund et al., 2003) and ISGRI (Lebrun et al., 2003) --- one of 
two detectors of the IBIS (Ubertini et al., 2003) $\gamma$-ray 
telescope. Both of these instruments use a principle of coding 
apperture to construct images of the sky in the field of view and
study individual sources.

JEM-X monitor is sensitive to photons in 3--35 keV energy 
band and has a field of view of 13\fdg2 in a diamiter (with only
4\fdg8 of it fully coded) and an angular resolution of 3\farcm35
(FWHM). A gas chamber with entrance window area of 490 cm$^2$ and
an energy resolution $\Delta E/E\sim16 @ 6keV$\%\ (FWHM) is used 
as a position-sensitive detector. The effective detector area for 
sources in the center of the field of view is just $\sim75$ cm$^2,$ 
for more then 80\%\ of the detector is shaded by opaque mask elements 
and collimator.

The ISGRI detector consists of 128$\times$128 CdTe 
semiconductor elements, with maximum sensitivity in a 18--200 keV 
energy band. Its energy resolution is $\Delta E/E\sim7$\% (FWHM). 
The total area of the detector is 2620 cm$^2$and the effective area
for sources in the center of the field of view is
$\sim 1100$ cm$^2$ (half of the detector is attenuated by opaque 
mask elements). The IBIS telescope has a field of view of $30\deg \times 
30 \deg$ ($9\deg\times 9\deg$ of it is fully coded) and an angular 
resolution of 12\arcmin\ (FWHM). Such a resolution allows one to
determine the position of bright sources with up to 2\arcmin\, accuracy.

An X-ray burst from AX J1754.2-2754 was registered on April 16, 
2005 --- during the deep (very large exposure) observation of 
the Galactic Center region by INTEGRAL observatory. But it was
discovered much later (see Chelovekov, Grebenev, 2007) --- as
one of the results of the project, dedicated to X-ray bursts 
search in open (for public access) archival IBIS/ISGRI telescope 
data of the INTEGRAL\footnote{While preparing the paper we found
out, that this burst caused an Integral Burst Announcement System
alert (event \#2463). IBAS scientists spread the messege 
(Merghetti et al., 2005) over GCN, reporting the event as
most likely not the actual GRB and pointing to a possible 
source of the observed activity --- AX J1754.2-2754. 
However, the properties and type of the burst as well as the nature 
of the source were not discussed in the report.} observatory. 
Data analysis methods,
used in this research are discribed in detail in paper by
Chelovekov et al. (2006). A later check revealed the burst 
under discussion in JEM-X telescope data as well.

\section*{BURST PROFILE}

\begin{figure}[]
\centering
\epsfig{file=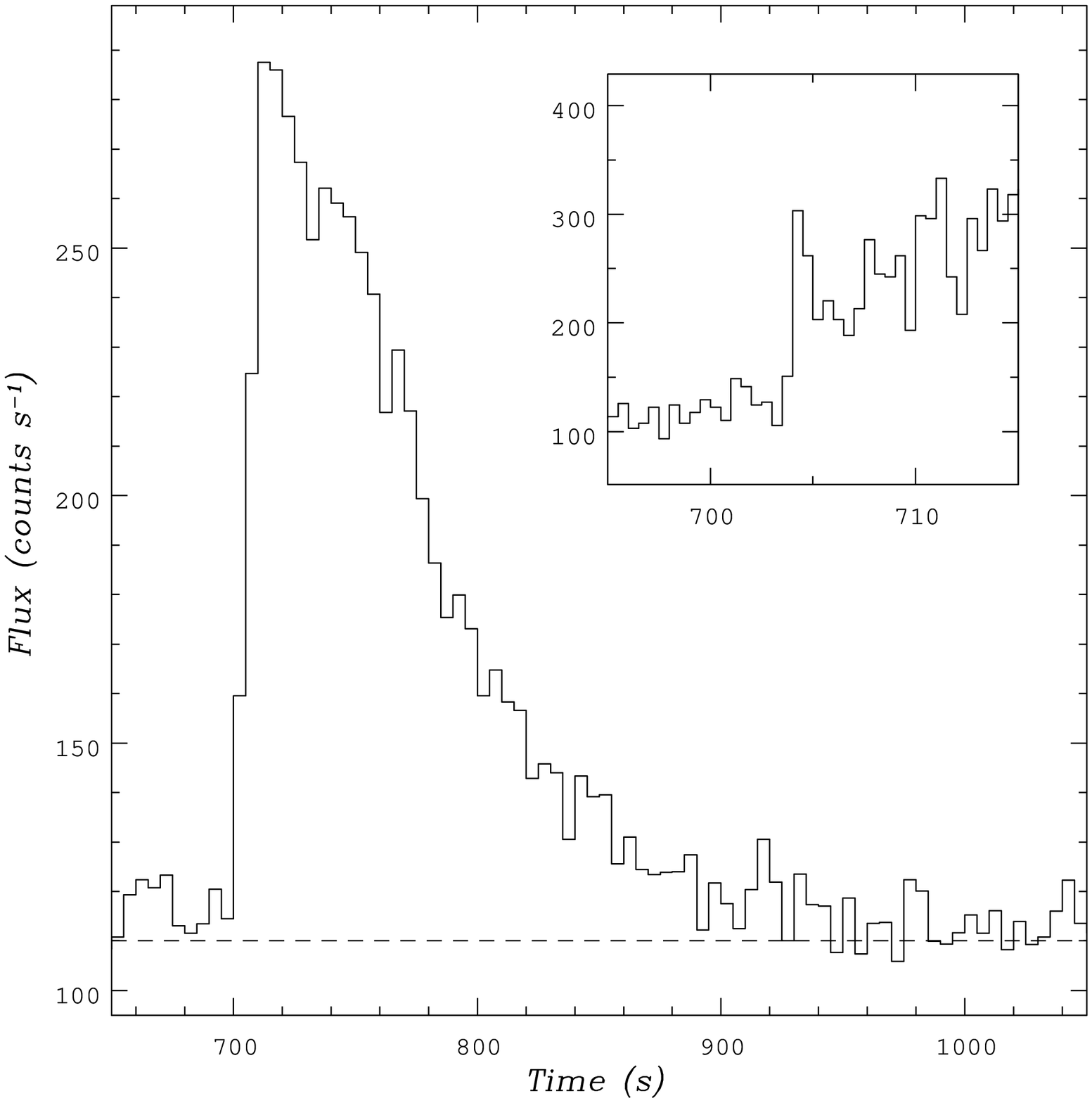,width=0.85\linewidth}

\caption{\rm A profile of an X-ray burst from AX\,J1754.2-2754, 
registered on April 16, 2005 by JEM-X telescope (3--20 keV energy band, 
è axis gives time in seconds from the begining of pointing 
UT 21\uh58\um35\us, time resolution is 5 s). The curve is corrected 
for a detector deadtime and a ``grey'' filter coefficient. The initial
phase of the burst with a smaller time resolution (1 s), to show its 
complex structure with a possible precursor, is presented in the inset.}
\label{fig:jemx_lc}
\end{figure}

Fig.\,{\ref{fig:jemx_lc}} shows JEM-X telescope lightcurve 
in 3--20 keV energy band during $\sim400$ s interval including 
the burst. A dashed line shows the preburst countrate level
due to persistant emission of the sources in the field of view
of the telescope as well as cosmic and instrumental backgrounds. 
Burst profile had a sharp (rise time $\sim10$ s) raise
and a long exponential decay (exponential decay time $t_{\rm e}=67\pm3$ 
s in the mentioned energy band). Such a profile is typical of type I
X-ray bursts (it is often refered to as FRED --- Fast Rise Exponential
Decay). Maximim countrate (corresponding to $\sim2$ Crab) was reached at UT
22\uh10\um25\us. An initial phase of the burst profile with a 
better time resolusion is shown in the inset. One can see
a fairly complex structure of the profile --- it is even possible to
assume the presence of a separate narrow (of $\la2$ s duration) peak
(precursor) $\sim 5-6$ s before the burst maximum.

\begin{figure}[p]
\centering
\epsfig{file=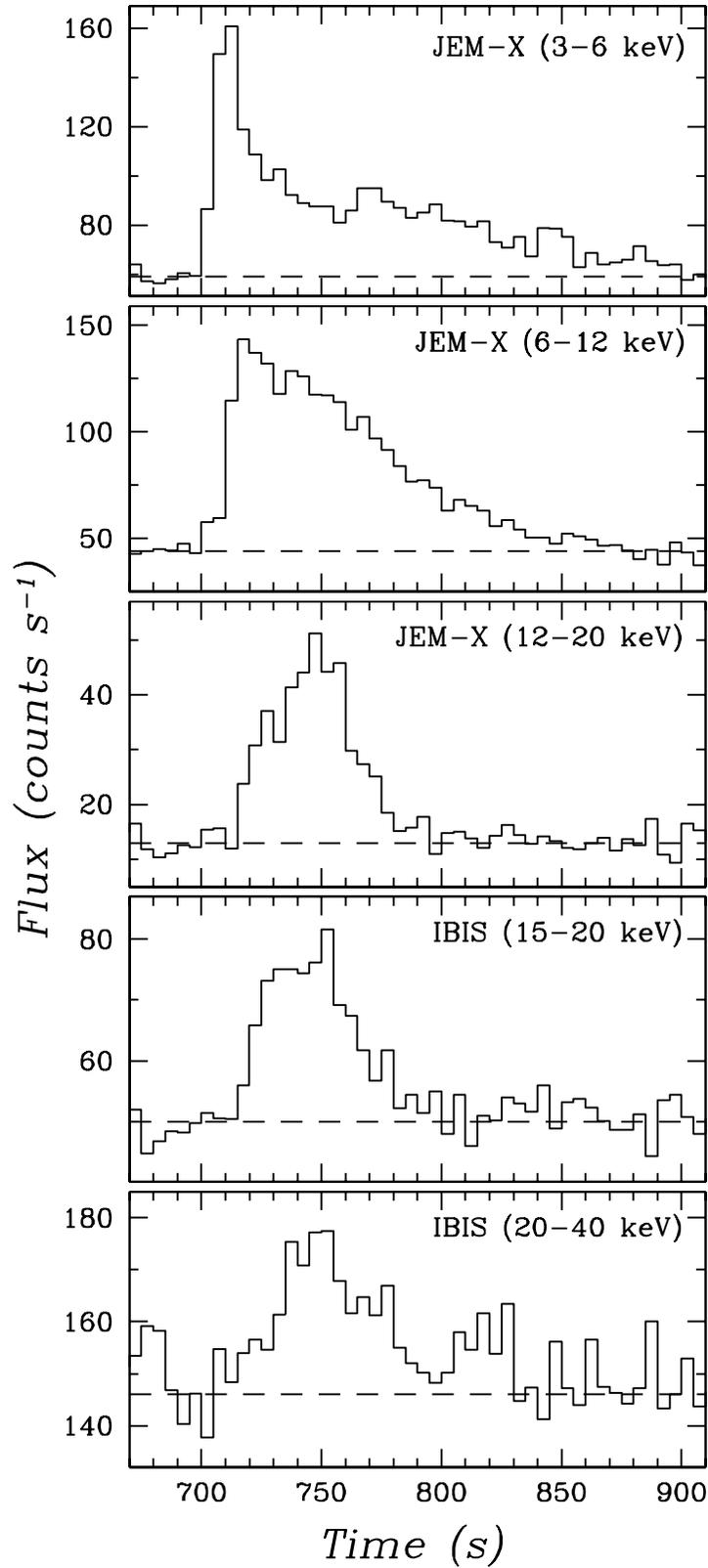,width=0.65\linewidth}

\caption{\rm A profile of an X-ray burst from AX\,J1754.2-2754, 
in various energy bands registered on April 16, 2005 by JEM-X 
and IBIS/ISGRI telescopes (è axis gives time in seconds from the 
begining of the observation UT 21\uh58\um35\us, time resolution 
is 5 s). All the curves are corrected for detector deadtime, all
JEM-X curves are also corrected for ``grey'' filter coefficient.}
\label{fig:jemxisgri_lc}
\end{figure}

It is of no lesser interest to study the burst profile with a 
better energy resolution and trace how its shape changes with 
energy. Fig. \,\ref{fig:jemxisgri_lc} (panels from top to bottom)
shows \mbox{JEM-X} burst profiles in 3--6, 6--12, 12--20 keV energy
bands and IBIS/ISGRI burst profiles in 15--20 and 20--40 keV energy
bands. While constructing all of these profiles as well as the profile
on fig.\,\ref{fig:jemx_lc} we used all the detector events with
no regard to the arrival direction of the origin photons.
All of these profiles are corrected for detector deadtime. All
JEM-X profiles are also corrected for ``grey'' filter, used in
case of large countrates (Lund et al., 2003).

Fig.\,\ref{fig:jemxisgri_lc} shows, that during initial ($\la20$
s) and final ($\ga80$ s) burst phases very soft radiation (3--6 keV)
is dominant in its profile. After reaching maximum in $\sim 10$ s
after the begining of the burst it started to decay rapidly with a 
charachteristic time $t_{\rm e}=31\pm6$ s, but then, after $\sim 30$ s, the 
decay flattened. The 6--12 keV emission reached its maximum in $\sim 20$ s
after the begining of the burst and later decayed with an exponential 
decay time $t_{\rm e}=115\pm6$ s. 
Hard (12--40 keV) emission appeared only for comperativly
short time interval (20--80 s) and reached its maximum around
UT 22\uh11\um04\us\ ($\sim 50$ s after the begining of the burst). 
Burst profile in this energy band had almost triangular shape.
All these features are typical of type I X-ray bursts, which are
associated with thermonuclear explosions of the matter, 
stored on the surface of the neutron star during the accretion 
in a binary system. The possible presence of the burst precursor and 
the registration of very soft radiation during the initial phase 
of the burst suggest that the neutron star atmosphere expansion 
due to Eddington critical luminosity of the source could take
place at this phase.

\section*{LOCALIZATION}

\begin{figure}[p]
\centering
\epsfig{file=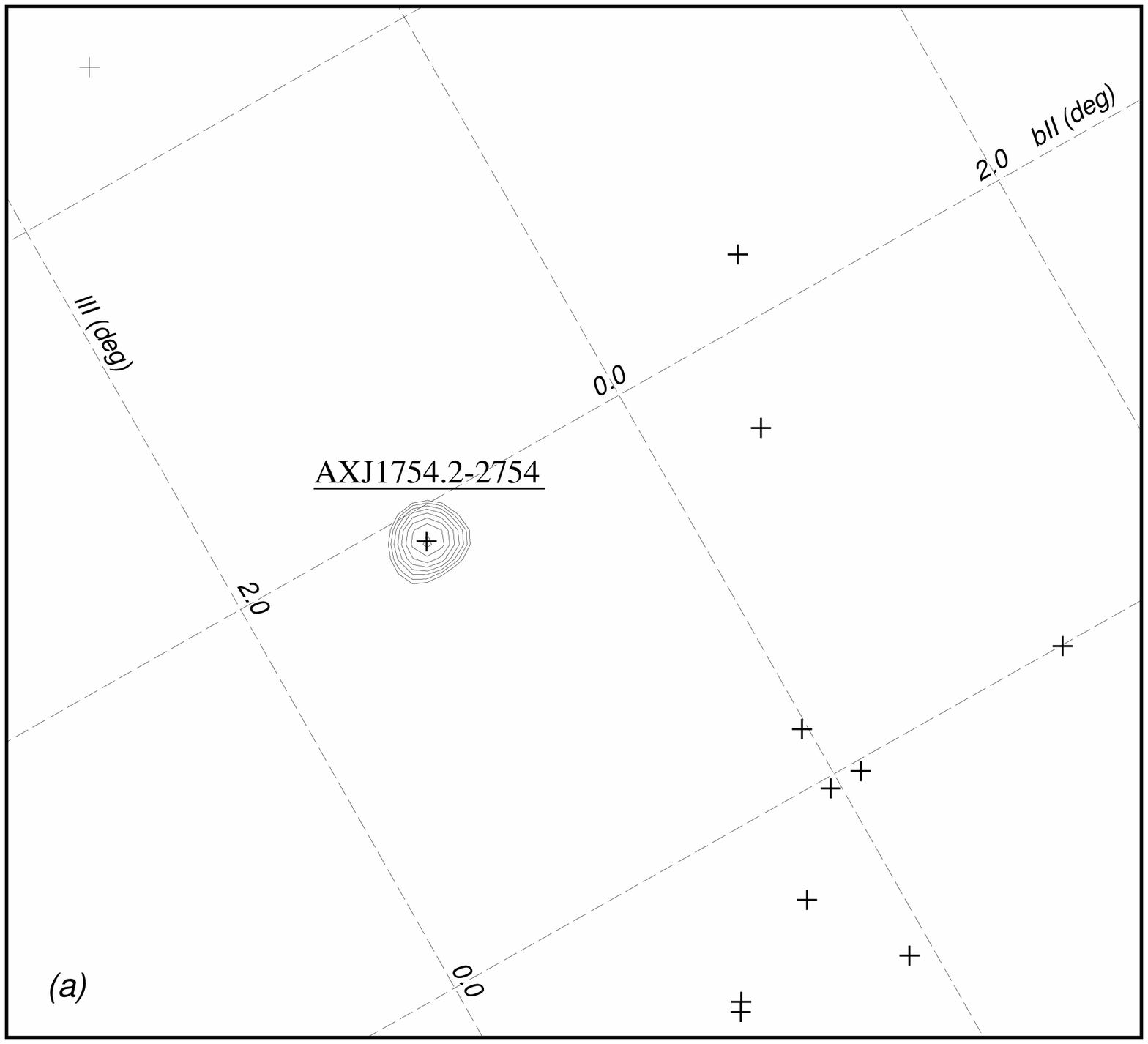,width=0.65\linewidth} 
\epsfig{file=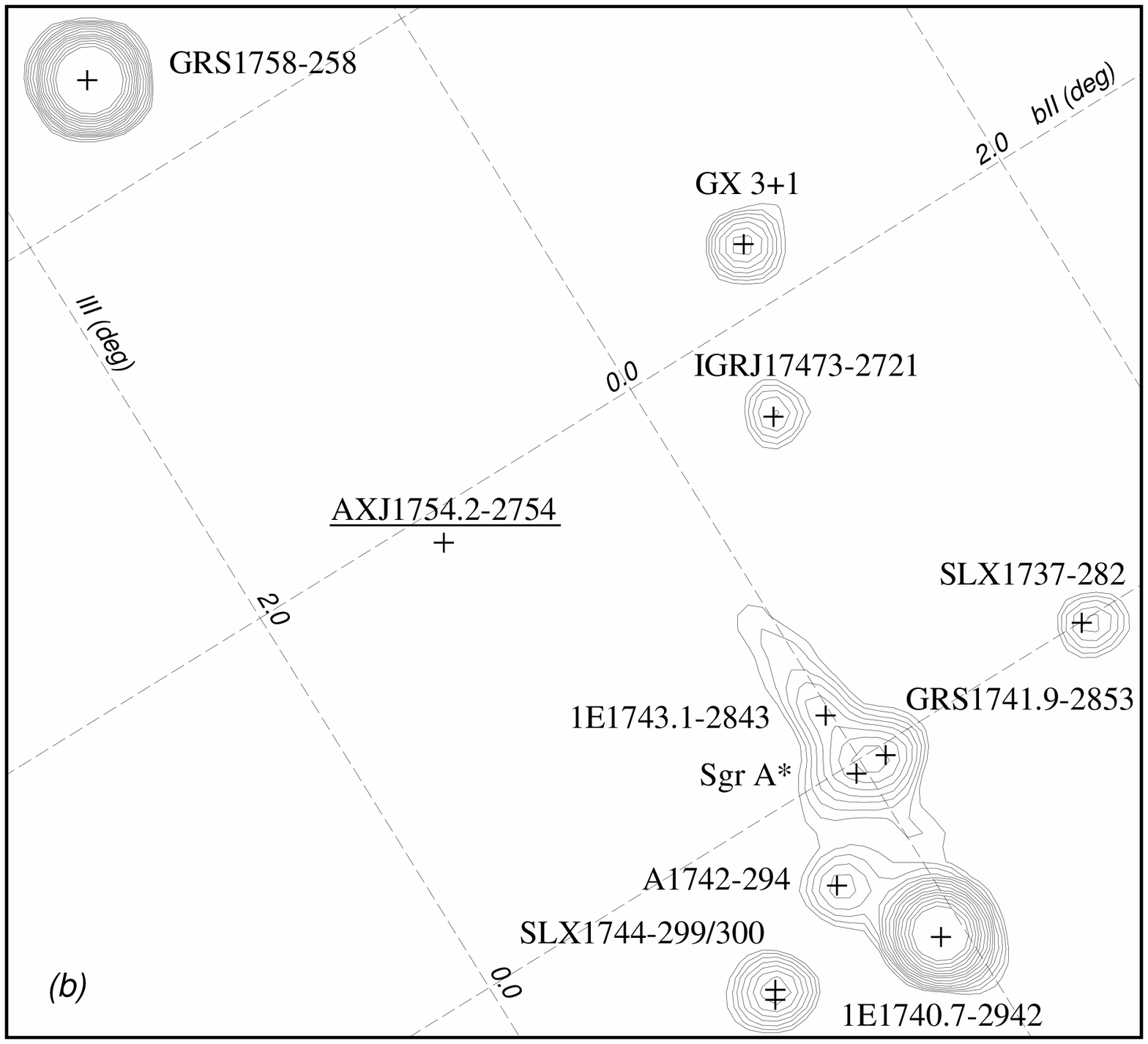,width=0.65\linewidth} 

\caption{\rm Images of a $2\fdg6\times2\fdg3$ sky region in the 
IBIS/ISGRI field of view: (a) during the X-ray burst,
registered from AX\,J1754.2-2754 (70 s exposure, 15--25 keV energy band),
and  (b) over the entire observing session (the entire orbit), except
the pointing, including the burst itself (201600 s exposure,
18--45 keV energy band). The contours show ereas of sources registration 
at signal-to-noise ratios $S/N=4.5, 5.4, 6.4, 7.7, 9.1, 11, 13, 16, ..., 45$.}
\label{fig:isgri_skyima}
\end{figure}

Fig.\,\ref{fig:isgri_skyima}a shows an image (map of the signal 
to noise ratio $S/N$), built based on IBIS/ISGRI data
in 15--25 keV energy band acumulated during first 70 s of the burst. 
Many sources including well known X-ray bursters such as GX\,3+1,
A\,1742-294, SLX\,1744-299/300, GRS\,1741.9-2853 (see 
fig.\,\ref{fig:isgri_skyima}b) are situtated within the field under 
consideration, but the image makes it clear, that none of them but
AX\,J1754.2-2754 was the source of the burst. It was the only
significant ($S/N\simeq16$, mean flux was $660\pm40$ mCrab) 
source detected during these 70 s in the field of view of the 
telescope.

\begin{figure}[p]
\centering
\epsfig{file=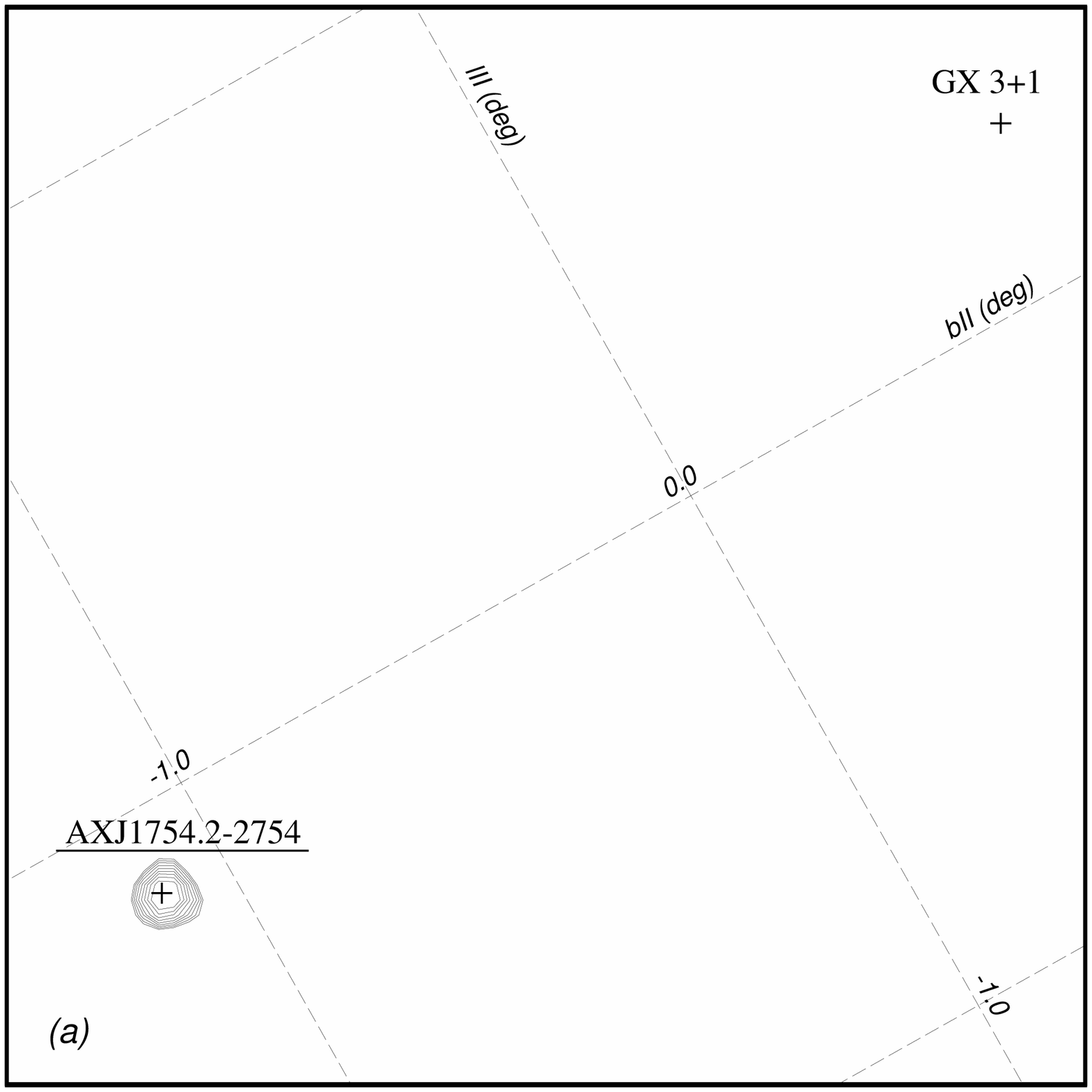,width=0.65\linewidth} 
\epsfig{file=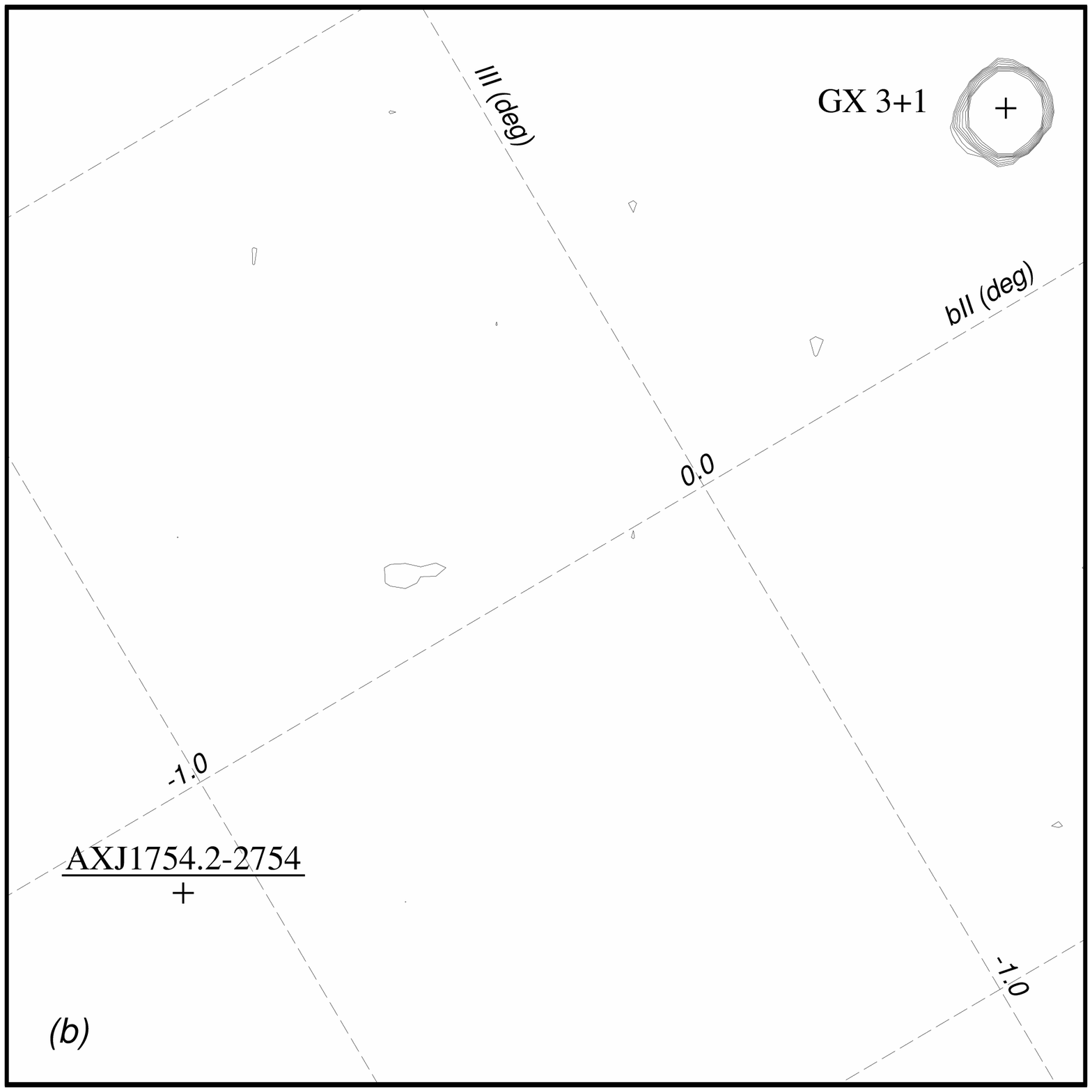,width=0.65\linewidth} 

\caption{\rm Images of a $1\fdg9\times1\fdg9$ sky region in the JEM-X field 
of view obtained in the 3-20 keV energy band: (a) during
the X-ray burst, registered from AX\,J1754.2-2754 (exposure 70 a),
and (b) over the entire observation session, except for the observation, 
including the burst itself, and some observations, 
during which the observatory was pointed more then $5\deg$ away from 
the source under consideration (exposure 97400 s). The contours show 
ereas of sources registration at signal-to-noise ratios
$S/N=4.5, 5.4, 6.4, 7.7, 9.1, 11, 13, 16, 19$.}
\label{fig:jemx_skyima}
\end{figure}

Fig. \,\ref{fig:jemx_skyima}a and b show simular, but smaller
3--20 keV JEM-X images. For this figure we specially used a field 
of the sky, including an X-ray burster GX\,3+1. Despite its
brightness (see fig.\,\ref{fig:jemx_skyima}b), this source was
not detected during 70 s of the burst by \mbox{JEM-X} telescope 
at significant level, unlike AX\,J1754.2-2754 ($S/N\simeq32$). 
At the same time, another very bright source GX\,5-1
($S/N\simeq22$) was registered in the field of view of the 
telescope, but it is not a burster. Besides, IBIS/ISGRI
data leaves no doubts that the burst came from the immediate
neighborhood of AX\,J1754.2-2754.

A more accurately determened position of the burst source accrording 
to JEM-X data $R.A.=17\uh54\um12\us$, $Decl.=-27\deg54\arcmin58\arcsec$
(epoch 2000.0, uncertainty $1\arcmin$), shows that it is 33\arcsec\ 
away from the position of AX\,J1754.2-2754, determened by the ASCA
satellite. Burst source position derived from IBIS/ISGRI data,
$R.A.=17\uh54\um13\us$, $Decl.=-27\deg54$\arcmin11\arcsec\
(uncertainty $2\arcmin$), is 41\arcsec\ away from this 
AX\,J1754.2-2754 position. So it is obvious, that precisely 
AX\,J1754.2-2754 was the source of the burst.


\section*{BURST SPECTRUM}

\begin{figure}[]
\centering
\epsfig{file=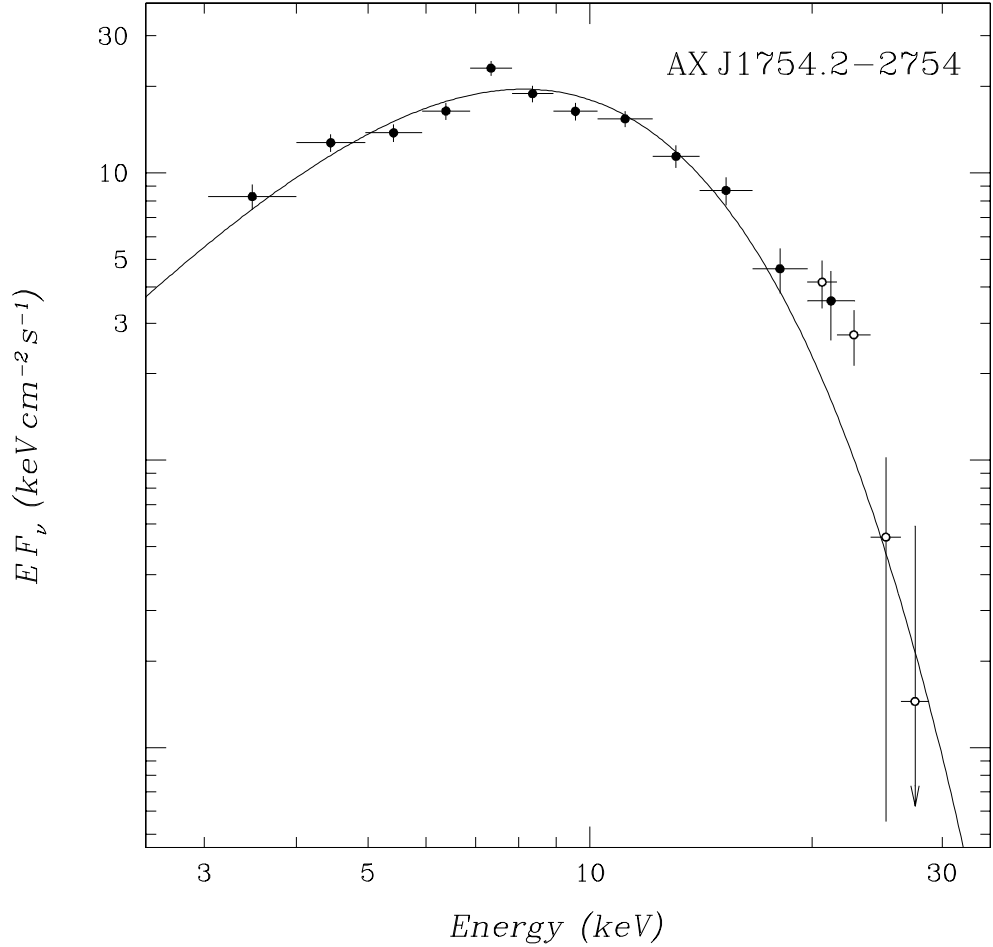,width=0.9\linewidth}
\caption{\rm Average spectrum of the X-ray burst from AX\,J1754.2-2754, 
obtained by JEM-X telescope (filled cirles) and IBIS/ISGRI telescopes
(empty cirles) during first 70 s of the burst. The solid line shows a 
black body model approximation of the spectrum.}
\label{fig:jemx_mean_70s_bspec}
\end{figure}

Fig.\,\ref{fig:jemx_mean_70s_bspec} shows the average spectrum of the
burst in $\nu F_{\nu}$ units, obtained by JEM-X (filled circles) 
and IBIS/ISGRI (empty circles) telescopes during its
first 70 s. A best spectrum approximation with black body model
is shown by solid line. We assumed that the interstellar 
absorbtion for this source corresponds to the hydrogen column density 
$N_{H}=2\times10^{22}$ cm$^{-2}$ for solar elemental abundance.
Parameters of the approximation ---
black body temperature $kT_{bb}=2.10\pm0.04$ keV and neutron star
photosphere radius $R_{bb}=10.3\pm2.9$ km --- are in a good agreement
with those measured for other bursters. At the same time
fig.\,\ref{fig:jemx_mean_70s_bspec} and the high value of $\chi^2=2.1$
normalized for number of deegrees of freedom for this approximation
show, that the fit is far from the ideal and this discripancy may 
well be expected. The burst spectrum may be strongly modified by
comptonization in the neutron star atmosphere, and even more then 
that: the average spectrum under discussion is a complicated combination
of black body spectra with different values of temperature, emmited 
at different stages of the burst.

\begin{figure}[]
\centering
\epsfig{file=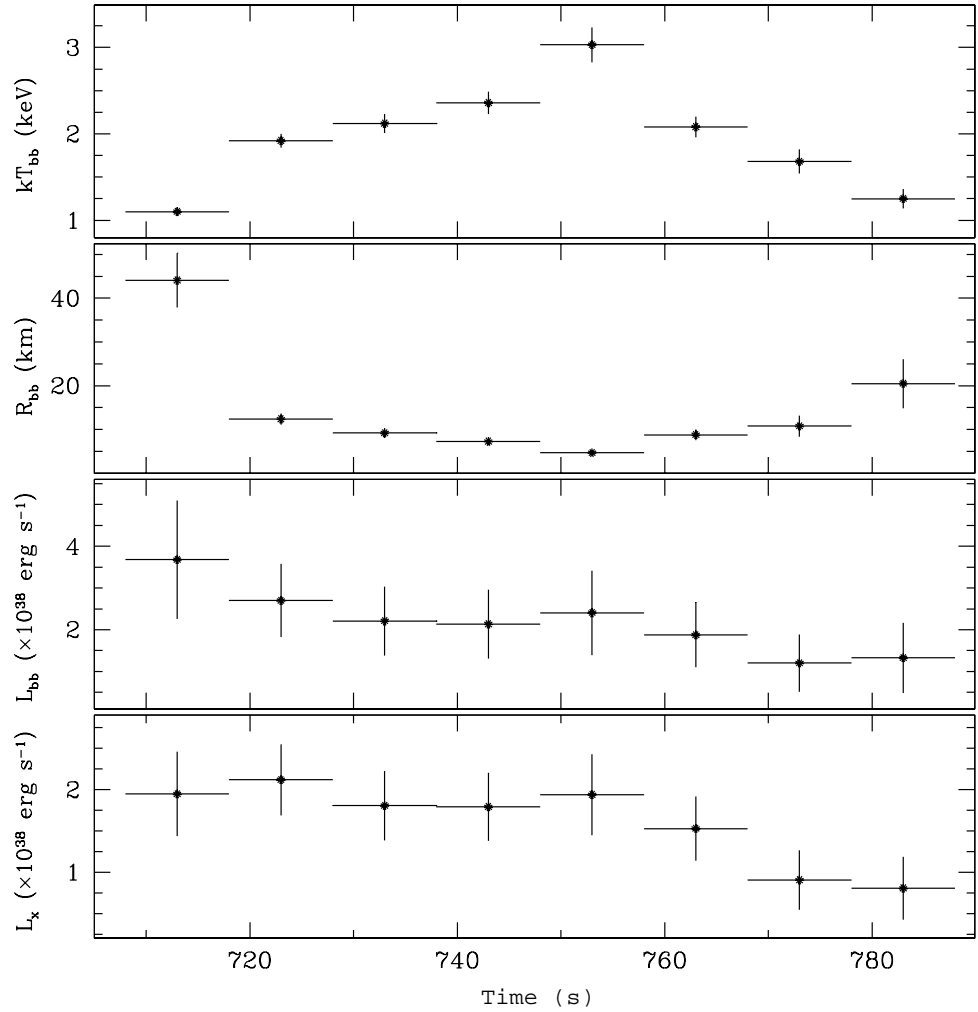,width=0.85\linewidth}
\caption{\rm Evolution of AX\,J1754.2-2754 spectrum parameters during 
the X-ray burst in a blackbody radiation model (from top to bottom:
evolution of temperature, radius of the emmiting body, bolometric 
luminosity, observed 3--20 keV luminosity.)}
\label{fig:jemx_spepar}
\end{figure}

To trace the evolution of the X-ray burst spectrum we constructed
and studied ``prompt'' spectra for eight successive time 
intervals of 10 s durration each. Fig.\,\ref{fig:jemx_spepar} presents
the results of this analysis. The spectrum for the interval, separated
by $\sim50$ s from the begining of the burst was the hardest as it
was clear even from the comparisson of burst profiles 
(fig.\,\ref{fig:jemxisgri_lc}). It is improtant to mention, that
estimated value of neutron star photosphere black body radius 
for the first 10 s interval was more then five times higher,
then the ones for other time intervals, while the black body
temperature of the photosphere was lowest ($\sim1$ keV) during 
this interval. It is obvious than, that a photosphic expansion 
took place during this interval due to luminosity reaching the 
Eddington limit value.

Fig.\,\ref{fig:isgri_skyima} and \ref{fig:jemx_skyima}
show, that the average AX\,J1754.2-2754 persistant X-ray flux over the
entire observing session, including the burst ($\sim3$ days), is
bellow the level of significant detection by JEM-X and IBIS/ISGRI
telescopes. A 3$\sigma$ upper limit for the 18-45 keV flux from 
the source in case of IBIS/ISGRI was 1.7 mCrab. 


\section*{DISCUSSION}

Our analysis shows, that X-ray burst, registered by INTEGRAL 
observatory telescopes on April 16, 2005 from AX\,J1754.2-2754
was of type I - a burst resulting from thermonuclear explosion 
on the surface of the neutron star. Registration of this burst
determines a nature of a compact object in this source --- it is 
a neutron star. Certain features of the initial phase of the burst 
profile are typical of neutron star photosphere expansion 
and cooling, which suggests, that luminosity of the source reached a 
critical Eddington limit
$L_{ed}\simeq2.8\times10^{38} (M/M_{\sun})/(1+X)$ erg s$^{-1}$, where
$M$ is a mass of neutron star and $X$ is a hidrogen aboundance in its 
atmosphere. Assuming the neutron star mass to be $1.4 M_{\sun}$, we
estimated the distance to AX\,J1754.2-2754
$$d=({L_{ed}}/{4\pi F_{max}})^{1/2}\simeq6.6\pm0.3\ \mbox{\rm kpc},$$
for a hidrogen atmosphere ($X=1$) and $d\simeq9.2\pm0.4$\ kpc for a
helium atmosphere ($X=0$).
Here $F_{max}=(3.84\pm0.15)\times 10^{-8}$ srg cm$^{-2}$ s$^{-1}$
is a maximum flux, registered during the burst. It is improtant to 
mention, that the vast bulk of burst energy is emitted in 3--20 keV energy 
band (see fig.\,\ref{fig:jemx_mean_70s_bspec}).


\section*{ACKNOWLEDGMENTS}

We wish to express our greatitude to S.V. Molkov for usefull discussions
and to M.G. Revnivtsev for an information of the corresponding IBAS event.
This research was based on INTEGRAL mission data, retrieved via the Russian 
and Europian Science Data Centers of this mission, and was carried out with 
support from Russian Foundation for Basic Research (project no. 05-02-17454), 
the Presidium of the Russian Academy of Sciences (the ``Origin and Evolution 
of Stars and Galaxies'' Program), and the Presidential Program for Support 
of Scientific Schools (proj no. NSh-1100.2006.2).


\clearpage



\end{document}